\documentclass[10pt,final,twocolumn,twoside]{IEEEtran}
\usepackage{graphicx} \usepackage{amsmath} \usepackage{amssymb}
\usepackage{color}
\usepackage{amssymb}
\usepackage{amsthm} 
\usepackage[linesnumbered,ruled,vlined]{algorithm2e}
\SetKwComment{Comment}{$\triangleright$\ }{}
\newlength\algowd
\def\savewd#1{\setbox0=\hbox{#1\hspace{.7in}}\algowd=\wd0\relax#1}
\newcommand\algolines[2]{\savewd{#1}%
  \Comment*{\parbox[t]{\dimexpr\algowidth-\algowd}{#2}}}
\usepackage{cite} 
\usepackage[latin1]{inputenc}
\usepackage[caption=false,font=footnotesize]{subfig} 
\usepackage{stfloats}
\usepackage{url}
\usepackage{mathtools}
\usepackage{varwidth}

      \theoremstyle{definition}      \theoremstyle{remark}      \theoremstyle{plain}     

           \newcommand{\mbf}[1]{\mathbf{#1}}     \newcommand{\mbs}[1]{\boldsymbol{#1}}

 \newcommand{\NF}{N_{\mbox{\scriptsize f}}}

  \newcommand{\ttoa}{\tau_{\mbox{\scriptsize toa}}}
  \newcommand{\httoa}{\hat{\tau}_{\mbox{\scriptsize toa}}}
  \newcommand{\ntoa}{n_{\mbox{\scriptsize toa}}}
  \newcommand{\hntoa}{\hat{n}_{\mbox{\scriptsize toa}}}
  \newcommand{\hrtoa}{\hat{\sigma}_{\mbox{\scriptsize toa}}^2}
  \newcommand{\vtoa}{v_{\mbox{\scriptsize toa}}}
  \newcommand{\NED}{N_{\mbox{\scriptsize ED}}}
  
  \newcommand{\NOBS}{N_{\mbox{\scriptsize obs}}}
  
  \newcommand{\Mprd}{\mbf{M}_{\mb{\scriptsize prd}}}
  \newcommand{\Minit}{\mbf{M}_{\mb{\scriptsize init}}}
  \newcommand{\sprd}{\mbf{s}_{\mb{\scriptsize prd}}}
  \newcommand{\sinit}{\mbf{s}_{\mb{\scriptsize init}}}

  \newcommand{\musn}{\mu_{H_1}}
  \newcommand{\rsn}{\sigma_{H_1}}
  \newcommand{\muno}{\mu_{H_0}}
  \newcommand{\rno}{\sigma_{H_0}}
  \newcommand{\vr}[1]{\sigma_{\mbox{\scriptsize{#1}}}}
  \newcommand{\xno}{x_{n_{\mbox{\scriptsize no}}}}
  \newcommand{\xsn}{x_{n_{\mbox{\scriptsize sn}}}}

  \newcommand{\wn}{\mbf{w}[n]}

 \newcommand{\WRX}{\omega_{\mbox{\scriptsize rx}}}

 \newcommand{\WTR}{\omega_{\mbox{\scriptsize tr}}}

 \newcommand{\TS}{T_{\mbox{\scriptsize s}}}

 \newcommand{\mb}[1]{\mbox{#1}}

\begin{document}

\title{Multi Detector Fusion of Dynamic TOA Estimation using Kalman Filter.}
\author{Vijaya Yajnanarayana, Satyam Dwivedi, Peter H\"{a}ndel}
\maketitle

% \date{\today} \author{Vijaya Yajnanarayana,   Satyam Dwivedi, Peter % Händel\\
%   ACCESS Linnaeus Center,\\
%   Signal Processing Lab,\\KTH Royal Institute of Technology,   % Stockholm, Sweden\\
%   email:\{vpy,dwivedi,ph\}@kth.se}
% %\author{Vijaya Yajnanarayana, Satyam Dwivedi, Peter H\"{a}ndel}
% \maketitle

\begin{abstract}
In this paper, we propose fusion of dynamic TOA (time of arrival) from multiple non-coherent detectors like energy detectors operating at sub-Nyquist rate through Kalman filtering. We also show that by using multiple of these energy detectors, we can achieve the performance of a digital matched filter implementation in the AWGN (additive white Gaussian noise) setting. We derive analytical expression for number of  energy detectors needed to achieve the matched filter performance. We demonstrate in simulation the validity of our analytical approach. Results indicate that number of energy detectors needed will be high at low SNRs and converge to a constant number as the SNR increases. We also study the performance of the strategy proposed using IEEE 802.15.4a CM1 channel model and show in simulation that two sub-Nyquist detectors are sufficient to match the performance of digital matched filter.  \\

\end{abstract}
%\vspace{-0.5in}
\section{Introduction}
\label{sec:intro}

An ultra wideband (UWB) system is based on spreading a low power signal in to large bandwidth. There are several UWB techniques, impulse radio based UWB (IR-UWB) schemes are most popular as they provide better performance and complexity trade-offs compared to other UWB schemes \cite{scholtz,moe_sch_2,Jinyun}. Narrow impulse signals that are used in IR-UWB schemes yield very fine time resolution and thus, can be used for accurate measurement of time of arrival (TOA) \cite{gezici-mag,poor-mag}.

Accurate TOA is essential in several applications including localization and communication. In localization, when the nodes
are synchronized, the TOA of the signal can be
used directly to obtain the range estimate. If nodes are not synchronized, TOA estimate is still needed for several ranging protocols to estimate the range. Localization information about the node can be derived from its range to anchor nodes \cite{poor-mag,monir-TOA}. In many IR-UWB communication systems, the message information is embedded in the location of the IR-UWB pulse. For example, pulse position modulation (PPM) and its variant modulation schemes. To demodulate the IR-UWB symbols in these modulation schemes, TOA estimation techniques can be used \cite{nikos-ppm,VJ-3,VJ-2}.

IR-UWB system with matched filter can provide optimal estimate of TOA for additive white Gaussian noise (AWGN) limited channels. The TOA is estimated by finding the peak location at which the filter attains the maximum value. The impulse response of the matched filter should closely approximate the transmit pulse shape. When implemented digitally, matched filter requires Nyquist-rate sampling \cite{cook-radar-book,turin-mf}. There are several digital UWB receiver architectures for which matched filter is suitable including \cite{Odonnell-Transciver,newaskar-2,newaskar-adc}. Digital receiver structure offers several benefits such as flexibility in design, reconfigurability and scalability \cite{moumita-book}. However, UWB signal occupy extremly large bandwidth and thus requires high-speed analog-to-digital converters (ADCs). These speeds demand the use of interleaved flash ADC \cite{newaskar-adc} or a bank of polyphase ADCs \cite{Odonnell-Transciver}. In addition, the ADC must support a large dynamic range to resolve the signal from the strong narrowband interferes. These aspects makes digital UWB architecture which operates at Nyquist-rate to be costly and power hungry \cite{newaskar-adc,paredes-csuwb,won-digitalUWB}.

Due to the above discussed practical concerns, energy detectors are an interesting alternative as they are less complex and can operate at sub-Nyquist rate \cite{oppermann-1,guvnec-2,oppermann-2}. However, these detectors suffer from the noise due to the square-law device and are sub-optimal in AWGN channels. Performance analysis of stand alone energy detector for TOA estimation in sub-Nyquist rate has been studied in \cite{gezici-book,Guvenc,guvnec-2}. However, how to achieve the performance of a digital matched filter using  multiple energy detectors for a dynamic TOA model is to the best of the authors knowledge unavailable. In this paper, we first show that TOA estimation using digital matched filter achieves Cram\'{e}r-Rao lower bound (CRLB) asymptotically with SNR. We show analytically that we can achieve the performance of a Nyquist rate matched filter, by using multiple sub-Nyquist rate energy detector estimates. We show that number of energy detectors needed to meet the matched filters performance is high at low SNRs and reduces as SNR increases, and finally converges to $4$ as SNR increases asymptotically. We propose joint fusion and tracking of dynamic TOA using multiple energy detector estimates through a suitable Kalman filter design. We assess the performance of fused TOA estimate in simulation for a dynamic TOA model from multiple energy detectors and compare it with the digital matched filter and demonstrate the validity of the claims.

This paper is organized as follows: In Section \ref{sec:m}, we discuss the signal and system model employed. In Section \ref{sec:Det}, we will discuss different types of detectors. Here, we will study the energy collection strategy for matched filter and energy detectors. In Section \ref{sec:AWGN}, performance analysis of these two detectors are studied for a static TOA case. Section \ref{sec:Kalman}, discusses the Kalman filter design for joint fusing and tracking of the energy detector estimates for a dynamic TOA model. Section \ref{sec:ss}, demonstrates the performance in simulation. Finally, Section \ref{sec:Conclusion} details the conclusions from the design and results demonstrated.

\begin{figure*}[!t]
  \centering
  \fbox{\includegraphics[scale=0.6]{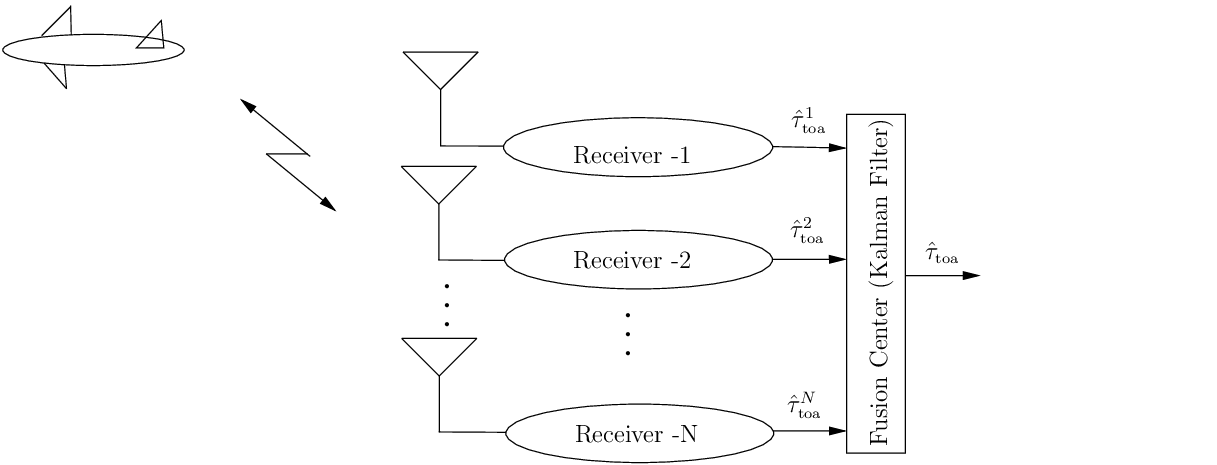}}
  \caption{Joint fusion and tracking structure using $N$ distributed receivers using low cost energy detectors operating at sub-Nyquist rate, tracking a dynamic TOA of a moving target.}
  \label{fig:KalmanStructure}
\end{figure*}

\section{Model}
\label{sec:m}

\subsection{Signal Model}
\label{sec:sm}

The signal model comprises of $\NF$ frames each having a unit energy pulse $s(t)$, given by

\begin{equation}
  \label{eq:tr}
  \WTR (t)=\sum\limits_{j=0}^{\NF}d_js(t-jT_f-c_jT_c),
\end{equation}
where each frame is of duration $T_f $ and the frame index is represented by $j$. The chip duration is represented by $T_c$ and $c_j \in \{0 \ldots N_c\}$ indicates the time-hopping code. $d_j\in \{\pm 1\}$ is the polarity code, which can be used along with time-hopping to smooth the signal spectrum. We consider the frame duration longer than the delay spread of the channel. 

A wide range of pulse shapes have been explored for UWB applications from rectangular to Gaussian. Gaussian and their derivatives, usually called monopulses, are effective due to the ease of construction and good resolution in both time and frequency. In many cost effective hardware designs, these shapes are generated without any dedicated circuits \cite{Ghavami2002,gavamibook,VJ-2,VJ-3}.  For analytical and simulation analysis, we have used the $2$nd order Gaussian pulse \cite{gezici-book},

\begin{equation}
  \label{eq:gauss_d2}
  s(t)=A\left(1-\frac{4\pi t^2}{\zeta^2}\right)\exp\left(\frac{-2\pi t^2}{\zeta^2}\right),
\end{equation}
The amplitude is adjusted through parameter, $A$, and pulse width is adjusted through parameter, $\zeta$.

The received signal is the distorted version of the transmit pulse with multipaths. The TOA is defined as the time elapsed for the first arrival path to reach the receiver from the transmitter. The received signal can be represented by 
\begin{equation}
  \label{eq:rx}
  \WRX (t)=\sum\limits_{j=0}^{\NF}d_jr(t-jT_f-c_jT_c) + n(t).
\end{equation}
where $r(t)=\sqrt{\frac{E_b}{\NF}}\sum\limits_{l=1}^{L}\alpha_l r_l(t-\tau_l)$, where $E_b$, is the captured energy and 
$\sum\limits_{l=0}^{l=L}\alpha_l^2=1$. The gain and received UWB pulse for the $l$-th tap is given by $\alpha_l$ and $r_l(t)$. The $n(t)$ is the AWGN process with zero mean and double-sided power spectral density of $N_0/2$. With out loss of generality, and for simplicity of analysis, we assume $c_j=0$ and $d_j=1$. TOA estimation problem is the estimate the first arraival path, $\tau_1=\ttoa$, in the received signal \eqref{eq:rx}.

\subsection{System Model}
\label{sec:sysm}

We define a system model as shown in Fig~\ref{fig:KalmanStructure}, where $N$ receivers each having an energy detector is employed to estimate TOA. When the target is moving then the estimated TOA is dynamic in nature. In these situations joint fusing and tracking using multiple sub-optimal estimates from the  sub-Nyquist receivers can yield better performance. We employ Kalman structure as shown in Fig.~\ref{fig:KalmanStructure} to jointly fuse and track the TOA. The multiple receivers can be distributed in nature and makes independent estimates of the TOA, which are then fused using a Kalman filter at the fusion center. All receivers should return the TOA for a reference location. Without loss of generality, we assume that the all receivers are equidistant from the target\footnote{If this condition is not met, then each receiver needs to appropriately scale the TOA based on the geometry of the reference location and actual location of the receiver.}. The proposed tracking and fusion strategy can also be employed in a single receiver architecture having a bank of energy detectors. 

\subsection{UWB Hardware}
\label{sec:sysm}
\begin{figure}[t]
  \centering
    \subfloat[Radio  module]{\includegraphics[width=10pc]{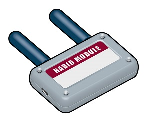}
     \label{fig:RadioModuleCartoon}}
    \subfloat[Printed circuit board (PCB)]{\includegraphics[width=10pc]{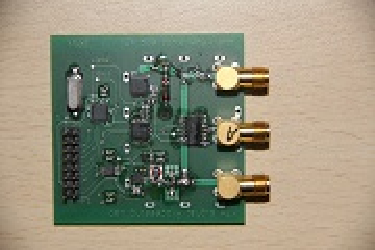}
    \label{fig:RadioModulePCB}}

  \caption{Iconic model of the in-house developed impulse radio UWB-platform of size $6\,\mbox{cm x }4\,\mbox{cm}$ for ranging and communication working in the $6\,\mb{GHz}$ regime with separate RX and TX antennas from Greenwavescientific. (details available at \cite{gw-antenna}) }
\label{fig:RadioModule}
\end{figure}

There are several low-complexity sub-Nyquist receivers based on energy detector available in literature \cite{gerosa,fontana,zhuo,Ales-2}. Figure \ref{fig:RadioModule} shows a graphical depiction of an in-house developed IR-UWB platform for ranging and communication. It uses a low cost pulse generator to generate sub-nano second pulses using step recovery diode (SRD), as described in \cite{Ales-exp}. The characterization and modeling of the UWB platform for a distance measurement system can be found in \cite{Ales-model}\cite{Ales-thr}. A detailed architectural description and experimental ranging results from a prototype of the platform have been published in \cite{Ales-2}. The power and range of the transceiver can be easily traded by controlling the amplitude, duty cycle and number of pulses per bit of transmission. Even though  proposed methods in this paper are not tied to a particular hardware, using the above discussed UWB hardware can yield significant performance benefits in terms of cost and power since they are implemented using low-complexity analog circuits.

\section{Detectors}
\label{sec:Det}
The two commonly employed receiver structures are as shown in the 
Fig.~\ref{fig:det}. The received signal passes through the low noise amplifier (LNA) and band pass filter (BPF) of bandwidth $B$. The output signal of BPF is converted in to energy samples, $\mbf{x}=\left(x_1,x_2,\ldots,x_{\NOBS}\right)$, from which TOA is estimated. Here, $\NOBS$, defines the number of energy samples used in the estimation. Based on the energy collection strategy different types of detectors exists. In this paper, we consider the matched filer which operates at Nyquist rate (1/$T_s$) and energy detectors operating at sub-Nyquist rate (1/$T_b$).

\subsection{Matched filter}
\begin{figure}[t]
  \centering
   \subfloat[Matched Filter]{\includegraphics[width=20pc]{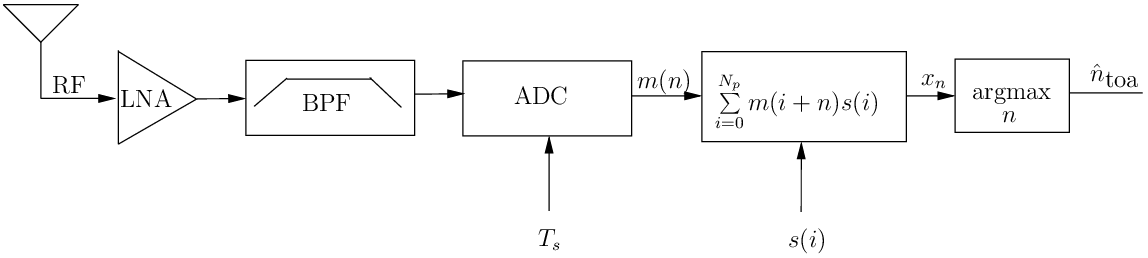}       \label{fig:MF}} \\
    \subfloat[Energy Detector]{\includegraphics[width=20pc]{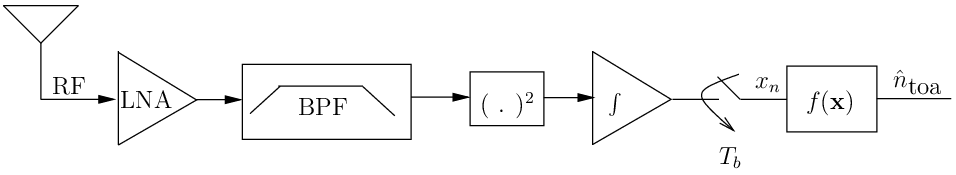}       \label{fig:ED}}
  \caption{Receiver structure with different detectors. The $s(i)$ and     $m(n)$ denotes the sampled version of transmit pulse and received     signal.}
  \label{fig:det}
\end{figure}

In  the matched filter, energy is collected by correlating the received samples with the transmit pulse shape as shown in Fig.~\ref{fig:MF}\cite{van2004detection}. Matched filter can be mathematically expressed as 
\begin{eqnarray}
  x_n&=&\sum\limits_{i=0}^{N_p}m(n+i)s(i), \label{eq:mf3} \\ 
  \hntoa&=&\underset{n}{\mbox{argmax}}(\mbf{x}),
\label{eq:mf4} \\
  \httoa&=&\hntoa T_s,\label{eq:mf5}
\end{eqnarray}
where, $s(i)=s(iT_s)$, represent the  digitized transmit pulse and $m(n)=\WRX(nT_s)$, represent the digitized received signal, sampled at interval $\TS$. In matched filter, the energy samples, $x_n$, are at Nyquist rate, $1/\TS$. Energy detectors, operating at the sub-Nyquist rates are an interesting alternative to digital matched filter.

\subsection{Energy Detector}
\label{subsec:ed}
 The structure for energy detector is as shown in Fig.~\ref{fig:ED}. The structure is amenable for a low-complexity analog implementation at sub-Nyquist rates \cite{oppermann-1,guvnec-2,oppermann-2}. In energy detector, the output signal from BPF is converted in to energy samples, $\mbf{x}=\left(x_1,x_2\ldots,x_{\NOBS}\right)$. TOA is estimated from the energy samples using the equation below
\begin{equation}
  \label{eq:ed1}
  \hntoa=f(\mbf{x}),
\end{equation}
\begin{equation}
  \label{eq:ed2}
 \httoa=\left(\hntoa-\frac{1}{2}\right)T_b,
\end{equation}
where, $f(\cdot)$, is the estimator function, which estimates the block-index/sample-index of the first arriving path. The function, $f(\cdot)$, is chosen based on the channel model. Equation \eqref{eq:ed2}, represents $\httoa$ as the mid-point of the corresponding estimated block, assuming that the true TOA is uniformly distributed with in this block.

\begin{figure}[t]
  \centering
  \includegraphics[scale=0.45]{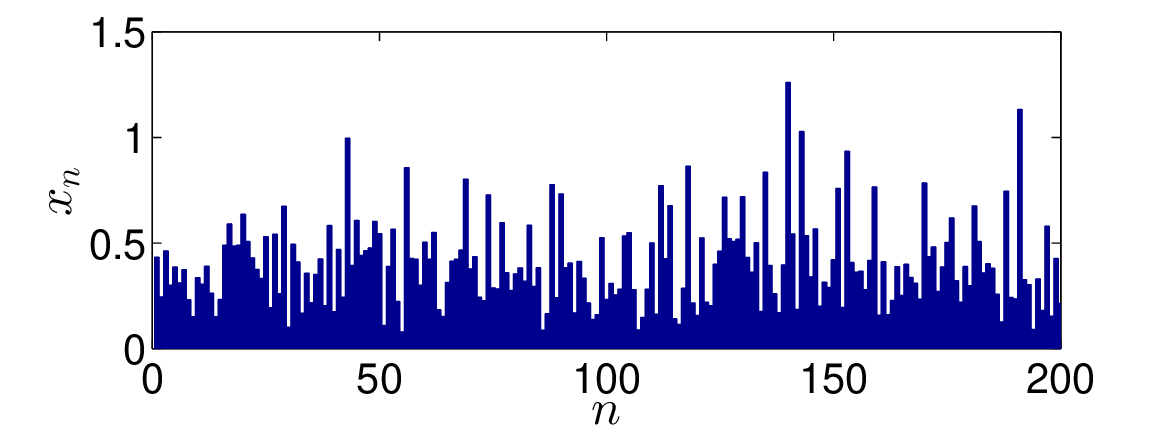}  
  \caption{Variation of energy samples, $x_n$, verses block index, $n$. Parameters $T_f=200~\mb{ns, }T_b=1~\mb{ns}, \mb{SNR}=0~\mbox{dB and  }\NOBS=200$ are considered.}
  \label{fig:bar}
\end{figure}

Without loss of generality, we consider the sub-Nyquist rate, $T_b=T_c$, in detector structure shown in Fig.~\ref{fig:det}. Therefore, each frame consists of, $\NOBS=T_f/T_c$, blocks, each with energy, $x_n\mb{, }n\in\left[1,\ldots,\NOBS\right]$, given by
\begin{equation}
  \label{eq:ed}
  x_n=\sum\limits_{i=0}^{\NF}\int_{(j-1)T_f+(n-1)T_b}^{(j-1)T_f+nT_b}|r(t)|^2dt.
\end{equation}
Typical variation of energy samples, $x_n$, verses $n$ is as shown in the Fig.~\ref{fig:bar}. In energy detector, the energy samples are at sub-Nyquist rate, $1/T_b$. 

\section{AWGN Channel Analysis}
\label{sec:AWGN}
The best performance in terms of mean-square-error (MSE) for an unbiased estimator is given by the CRLB and for a TOA estimation problem this is given by \cite{gezici-mag,dardari}:
\begin{equation}
  \label{eq:crb}
  \sigma_{\tau}^2\ge \frac{1}{8\pi^2\mb{SNR}\beta^2},    
\end{equation}
where $\beta$, is the effective signal bandwidth defined by
\begin{equation}
  \label{eq:beta}
  \beta^2=\left[ \frac{\int_{-\infty}^{\infty}f^2|S(f)|^2df}{\int_{-\infty}^{\infty}|S(f)|^2.df}   \right],
\end{equation}
where $S(f)$, is the Fourier transform of the transmit pulse, $s(t)$.

In AWGN channel model, we consider a single-path model, with $L=1$, $\alpha_1=1$ and $\tau_1=\ttoa$ in \eqref{eq:rx}. The probability density of the $x_n$, for matched filter and energy detector depends on whether the particular energy sample, $x_n$, is signal+noise sample or  noise-only sample. For matched filter it can be shown that the probability density functions under hypothesis, $H_1\cong x_n$ is signal+noise sample, and
 $H_0\cong x_n$ is noise-only sample, is given by 
 \begin{equation}
   \label{eq:pdf_mf}
   p(x_n)\sim
   \begin{cases}
      \mathcal{N}(\sqrt{E_b},\sigma^2)& \text{under } H_1,\\
      \mathcal{N}(0,\sigma^2)& \text{under } H_0,\\
   \end{cases}
 \end{equation}
where $\sigma^2=N_0/2$, is the variance of the noise samples and $\mathcal{N}$ denotes the Gaussian distribution. 

To derive the  probability distribution function (PDF) of the energy samples in energy detectors case, the function $f(\cdot)$ in \eqref{eq:ed1} need to be defined. 
Under the AWGN assumption, the received signal will have a single path, whose delay represents the TOA. Thus, optimal TOA estimation strategy here would be to pick the energy sample having maximum energy as $\hntoa$, this selection criteria is called maximal energy selection (MES) and is given by \cite{gezici-book} 

\begin{eqnarray}  
  \hntoa&=&\max(\mbf{x}), \label{eq:eda1} \\
  \httoa&=&\left(\hntoa-\frac{1}{2}\right)T_b, \label{eq:eda2}
\end{eqnarray}
The complementary cumulative distribution function (CCDF) of energy samples, $x_n$ for energy detector, is given by \cite{dardari,dardari-3,sahinoglu,gezici-book}

\begin{equation}
  \label{eq:eda3}
P(x_n>\eta)=
\begin{cases}
Q_{M'/2}\left(\frac{E_b}{\sigma},\frac{\sqrt{\eta}}{\sigma}\right) & \text{under } H_1 \\
\exp\left(-\frac{\eta\NF}{N_o}\right)\sum\limits_{i=0}^{M'/2-1}\frac{1}{i!}\left(\frac{\eta\NF}{N_0}\right)^i & \text{under } H_0 
\end{cases}
\end{equation}
where $Q_Z(a,b)$ denotes the Marcum-Q-function with parameter $Z$, $P$ denotes the probability and $M'\approx\NF(2BT_b+1)$, denotes the degrees of freedom (DOF). At high SNRs the above equation can be approximated to \cite{gezici-book}
\begin{equation}
  \label{eq:eda4}
   p(x_n)\sim
  \begin{cases}
      \mathcal{N}(\musn,\rsn^2)& \text{under } H_1,\\
      \mathcal{N}(\muno,\rno^2)& \text{under } H_0,\\   
  \end{cases}
\end{equation}

The mean and variance of $x_n$ under both hypotheses is given by
\begin{eqnarray}
  \musn&=&\NF M\sigma^2+E_b,   \label{eq:eda5}\\
  \rsn^2&=&2 \NF M\sigma^4+4\sigma^2E_b^2, \label{eq:eda6}\\
  \muno&=&\NF M\sigma^2,   \label{eq:eda7}\\
  \rno^2&=&2 \NF M\sigma^4, \label{eq:eda8}
\end{eqnarray}
where $M=2BT_b+1$, is the degrees of freedom of noise and $\sigma^2=N_0/2$.

If we choose the MES criteria for TOA estimation for both matched filters and energy detectors as given in \eqref{eq:mf3}~-~\eqref{eq:mf5} and \eqref{eq:eda1}~-~\eqref{eq:eda2} respectively, then an error can occur, if any one of the noise-only energy sample is higher than the  signal+noise energy sample. That is
\begin{equation}
  \label{eq:est1}
  \xno>\xsn,  
\end{equation}
where $\xsn$ and $\xno$ are energies of signal+noise and noise only sample. For AWGN channel, there will be only one signal+noise energy sample and all other samples are noise-only. The correct selection will happen when
$x_{n}<x_{sn}$, for all the energy samples except for the signal+noise energy sample. Thus, the probabilty of correct selection, $P_s$, and  probalility of error selection, $P_e$, is given by  \cite{sahinoglu}

\begin{eqnarray}
  \label{eq:est2}
  P_s&=&\int\limits_{x_n=0}^{\infty}\left(1-\mbf{Q}\left(\frac{x_n-\mu_{ \mbox{\scriptsize no}}}{\sigma_{\mbox{\scriptsize no}}}\right)\right)^{\NOBS-1}p(x_n)dx_n,\nonumber \\
  P_e&=&1-P_s,
\end{eqnarray}
where  $p(x_n)$ is the density function of the energy samples for the detectors under hypothesis $H_1$. $\mbf{Q}$ denotes the CCDF function of $x_n$. Here $\mu_{ \mbox{\scriptsize no}}$, denotes the mean values of $x_n$ for matched filter and energy detectors and is given by  0 and $\NF M\sigma^2$  respectively. $\sigma_{ \mbox{\scriptsize no}}^2$, denotes the variance of $x_n$ for matched filter and energy detectors and is given by  $\sigma^2$ and $2 \NF M\sigma^4$  respectively.

\begin{figure}[t]
  \centering
  \includegraphics[width=18pc]{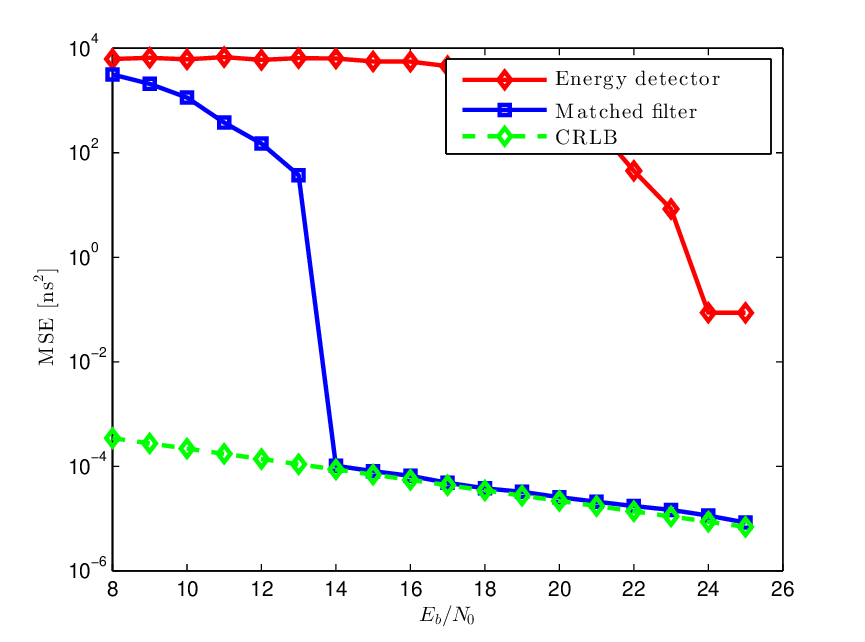}
  \caption{Variation of MSE with SNR for energy detector and matched     filter. MSE is evaluated by averaging the estimated TOA for $1000$ random $\ttoa$ drawn from $\mathcal{U}[0,T_f=200~\mb{ns}]$. Matched filter asymptotically reaches the CRLB bound.}
  \label{fig:MSE_Vs_SNR}
\end{figure}
\begin{algorithm}[t]
\DontPrintSemicolon 
\caption{Kalman filter for tracking and fusing of TOA from multiple sub-Nyquist energy detector estimates}
\label{kalman}
\small
\KwIn{Prior state, $\sinit$, prior state's covariance, $\Minit$, and covariance of plant noise, $\mathcal{Q}$. }
\KwOut{Tracked and fused TOA estimates, $\mbf{\httoa}$ and variance of estimate, $\hrtoa$}
$\mbf{s}[1] \gets \sinit$ \Comment*[r]{Initial state}
$\mbf{M}[1] \gets \Minit$ \Comment*[r]{Initial states Covariance} 
$\httoa[1] \gets \sinit[1]$ \Comment*[r]{Initial TOA estimate}  
$\hrtoa[1] \gets \Minit[1](1,1)$ \Comment*[r]{Initial MSE estimate}  
\For{$i \gets 1$ \textbf{to} $n$} {
  $\sprd \gets \mbf{A}\mbf{s}[i]$ \Comment*[r]{Pridicted state} 
  $\Mprd \gets \mbf{A}\mbf{M}[i]\mbf{A}'+\mbf{\mathcal{Q}}$   \Comment*[r]{Predicted MSE}
  \algolines{}{Compute Kalman gain}
  $\mbf{K}=(\Mprd \mbf{H}^{\scriptsize \mb{T}})(\mbf{C} +     \mbf{H}
    \Mprd \mbf{H}^{\scriptsize \mb{T}})^{\scriptsize -1}$\;   
    \algolines{}{Update state from observations ($\mbf{y}$[n])}
  $\mbf{s}[i+1]=\sprd+\mbf{K}(\mbf{y}[i]-\mbf{H}\sprd)$ \;
  $\httoa[i+1]=\mbf{s}[i+1](1)$ \Comment*[r]{Save fused TOA ($\httoa$)}
  $\mbf{M[i+1]}=(\mbf{I}_2- \mbf{K}\mbf{H})\Mprd$ \Comment*[r]{Update MSE} 
 $\hrtoa[i+1]=\mbf{M[i+1]}(1,1)$\Comment*[r]{Save MSE}
}
\Return{$\mbf{\hat{\tau}}_{\mbox{\scriptsize toa}},\mbf{\hrtoa} $}\;
\end{algorithm}

\begin{figure*}[!b]
\centering
\fbox{
\subfloat[Observed TOA by the energy detectors]{\includegraphics[scale=0.53]{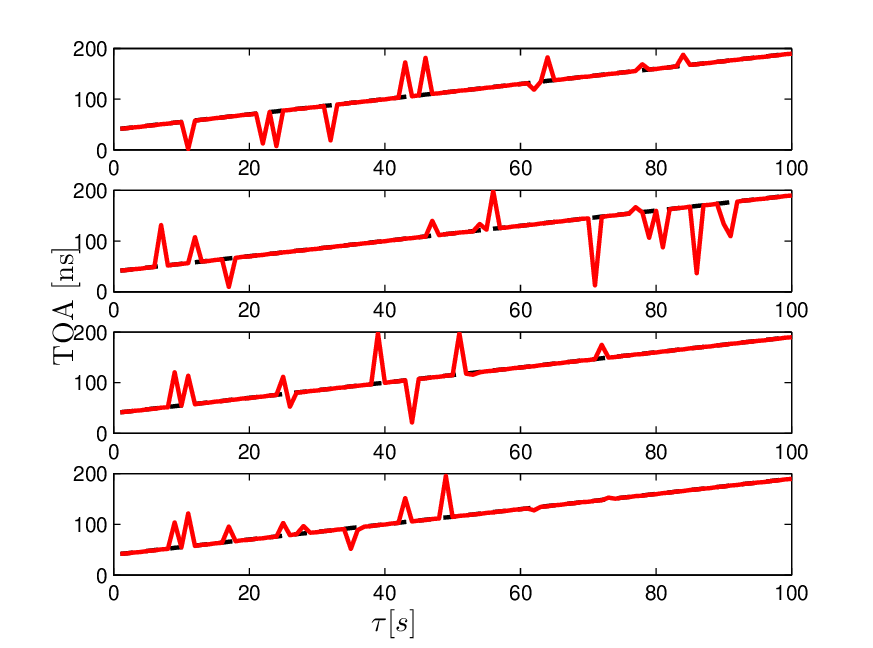}                 \label{fig:4_detectors_observations_12dB}}
\subfloat[OR fusion]{\includegraphics[scale=0.53]{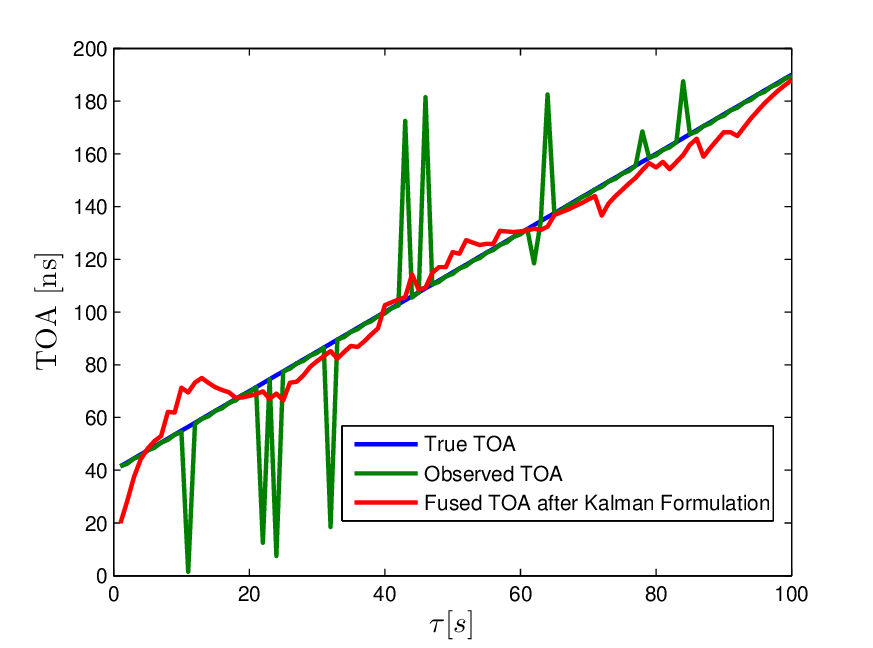} 
  \label{fig:4_detectors_tracked_12dB}}
}
\caption{Fusion of multiple Detectors using Kalman Formulation.}
\label{fig:fusion}
\end{figure*}

It is not straight forward to arrive at the closed form expression  for energy detector's $P_s$ and $P_e$ due to complex distribution functions \eqref{eq:eda3}. There is however a loss of performance in energy detectors due to squaring and integration operation at sub-Nyquist rates compared to matched filter, operating at Nyquist rate. Figure \ref{fig:MSE_Vs_SNR}, shows the mean square error (MSE), performance for matched filter and energy detector. MSE is  evaluated by averaging the estimated TOA for $1000$ random TOAs ($\ttoa$) drawn from $\mathcal{U}[0,T_f=200~\mb{ns}]$. The sampling rate ($1/T_s$) used by matched filter is $1000~\mb{GHz}$\footnote{Since TOA is continuous in time, very high sampling rate is employed to demonstrate that matched filter will indeed reach the CRLB bound without 
much ambiguity due to discretization of time due to sampling.} and sampling rate ($1/T_b$) used by energy detector is $1/1000$-th that of matched filter and is equal to $1~\mb{GHz}$.

At high SNRs, to achieve the performance of the matched filter using multiple energy detectors, we need the covariance of the signal+noise energy sample (under hypothesis $H_1$) to be same. We can approximate the probability distribution of energy samples for energy detectors in to a Gaussian distribution at high SNR as shown in \eqref{eq:eda4}. If there are  $K$ receivers having energy detectors to independently estimate TOA, then signal+noise block in each detector will have a probability distribution given by

\begin{equation}
  \label{eq:ei}
  e_i\sim \mathcal{N}\left( \musn,\rsn^2\right), i\in[1,\cdots,K].
\end{equation}
Since $e_i$s are independent and identically distributed, then the best estimate for the signal+noise block using $K$ energy detectors is given by \cite{kay-est-book} ,
\begin{equation}
  \label{eq:avg}
  e=\frac{1}{\NED}\sum\limits_{k=1}^{K}e_k.
\end{equation}
The probability distribution function of $e$ is given by\footnote{See \cite{pop-book} to obtain the probability distribution function of a function of Gaussian random variable.}
\begin{equation}
  \label{eq:pdfe}
  e\sim\mathcal{N}\left(\musn,\frac{\rsn^2}{K}\right).
\end{equation}
Let $\vr{MF}^2=\sigma^2$ and $\vr{ED}^2=\rsn^2$ denote the variance of the $x_n$ under $H_1$ for matched filter and energy detector. From \eqref{eq:pdf_mf} and \eqref{eq:pdfe} to achieve the performance of the matched filter using multiple energy detectors, we need 
\begin{equation}
  \label{eq:neds}
  \vr{MF}^2=\frac{1}{N_{\mbox{\scriptsize ED}}}\sigma_{\mbox{\scriptsize{ED}}}^2,
\end{equation}
where, $\NED$, is the number of energy detectors needed to have the same performance as that of matched filter. For normalized energy per bit ($E_b=1$), and using the variance for signal+noise sample from \eqref{eq:pdf_mf}, \eqref{eq:eda4} and \eqref{eq:eda6} we get
\begin{eqnarray}
  \NED&=&\lim_{\mbox{SNR} \to \infty}\frac{\sigma_{\mbox{\scriptsize{ED}}}^2}{\sigma_{\mbox{\scriptsize{MF}}}^2}\\
  \NED&=&\lim_{\sigma^2 \to 0}\frac{\left({2M\sigma^4+4\sigma^2}\right)}{\sigma^2}, \label{eq:asym1}\\
   &=&4. \label{eq:neds2} 
\end{eqnarray}
Thus, from \eqref{eq:neds2}, asymptotically, with the increase of SNR ($E_b/N_0$), the number of energy detectors needed to achieve the same performance as digital matched filter is equal to $4$. We will show in  simulation that this phenomenon is indeed true in the later section.

\section{Multipath Channel Analysis}
\label{sec:multi}
\begin{figure}[t]
  \centering
  \includegraphics[width=18pc]{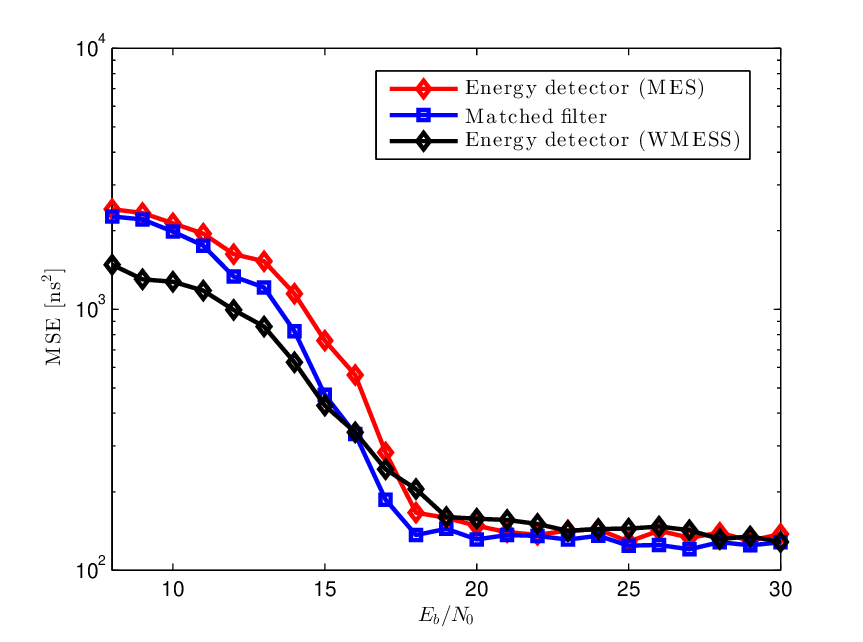}
  \caption{Variation of MSE with SNR for energy detector (with selection criteria MES and WMESS) and matched  filter. MSE is evaluated by averaging the estimated TOA for $1000$ random $\ttoa$ drawn from $\mathcal{U}[0,100~\mb{ns}]$.}
  \label{fig:MSE_Vs_SNR_multi}
\end{figure}

Many UWB ranging applications have channel response with several multipath components, i.e. the received pulse in \eqref{eq:rx} has $(\alpha_1,\alpha_2\ldots \alpha_L;\tau_1,\tau_2\ldots,\tau_L)$, where $L$ is the number of multipaths. The TOA estimation problem is to identify the leading edge (first arriving path, $\tau_1$). In multipath UWB channels, the matched filtering performance is not optimal since the shape of the pulse is lost in the channel due to the frequency selective fading. Also, the magnitude of the energy sample containing first arriving path may be smaller than the peak energy sample, therefore, using MES criteria for TOA estimation, as represented  in \eqref{eq:eda1} and \eqref{eq:mf4} does not always yeild true TOA  \cite{uwb-channel-model}. Figure \ref{fig:MSE_Vs_SNR_multi}, shows the  performance in terms of MSE for matched filter and energy detector for 802.15.4a, residential LOS channel (CM1 model) \cite{uwb-channel-model}. MSE is  evaluated by averaging the estimated TOA for $1000$ random $\ttoa$ drawn from $\mathcal{U}[0,100~\mb{ns}]$. The sampling rate ($1/T_s$) used by matched filter is $8~\mb{GHz}$ and sampling rate ($1/T_b$) used by energy detector is $1/8$-th that of matched filter and is equal to $1~\mb{GHz}$. 

\begin{figure}[t]
  \centering
  \includegraphics[width=18pc]{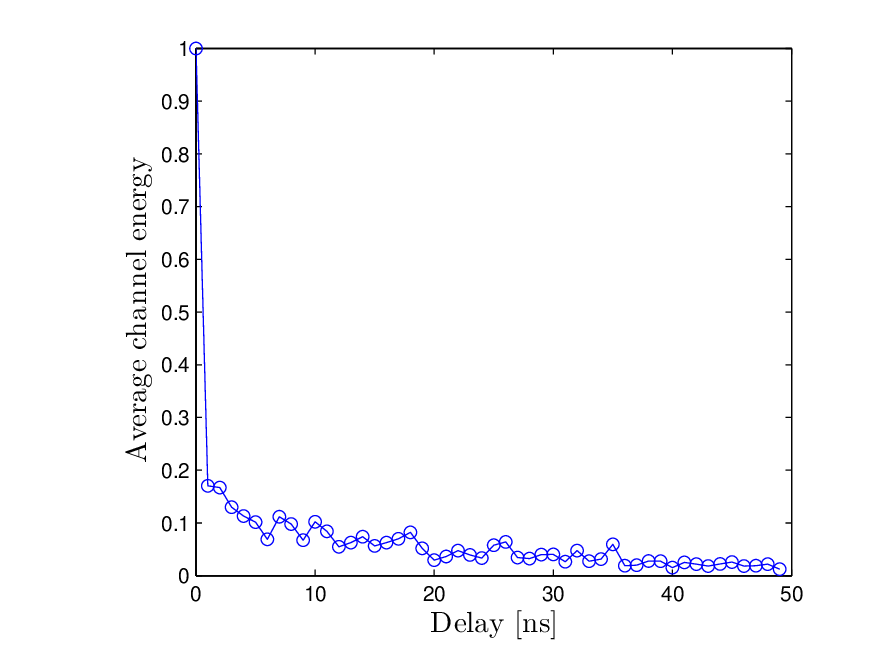}
   \caption{Averaged Channel energy profile for 802.15.4a CM1 channel model. 100 different channel realizations are averaged.}
  \label{fig:PDP}
\end{figure}

The performance of the energy detectors can be improved by considering the a-priori information such as power delay profile (PDP). To accomplish this, we can repose the TOA estimation problem as a multiple hypothesis testing problem. The energy samples vector, $\mbf{x}$, is a $1 \times \NOBS$ vector, out of which, $N_e$ blocks are having multipath signals, $N_eT_b$, represents the excess delay of the channel. If we define $H_k$, as the hypothesis that the k-th sample denotes the first path arrival, then $H_k= \mbf{x}$, with,

\begin{equation}
  \label{eq:multi-h}
  x(n)=
  \begin{cases}
    \begin{split}
      &\int_{(n-1)T_b}^{nT_b}n^2(t).dt,\\
      &\qquad \mbox{for } n=1,\ldots,k-1  \\
      &\int_{(n-1)T_b}^{nT_b}|\WRX(t)+n(t)|^2.dt,\\ 
      &\qquad \mbox{for }n=k,\ldots,k+N_e-1  \\
      &\int_{(n-1)T_b}^{nT_b}n^2(t).dt, \\
      &\qquad \mbox{for }n=k+N_e,\ldots,\NOBS.
    \end{split}
  \end{cases}
\end{equation}
Here, $x(n)$, denotes the $n$-th element of $\mbf{x}$ and $H_{\ntoa}$, is the true hypothesis \cite{gezici-book}. If the PDP for the channel is available, then we can modify the hypothesis test from MES to weighted maximum energy sum selection (WMESS), and is given by 
\begin{equation}
  \label{eq:wmess}
  \hntoa=\arg \underset{k \in [1,\ldots,\NOBS ]}{\max}\left\{x(k:k+N_e)\mbs{\mathcal{E}}\right\},
\end{equation}
where, $x(k:k+N_e)$ is a $1\times N_e$ vector having the $N_e$ elements from $\mbf{x}$ starting from $k$ and $\mbs{\mathcal{E}}$ is the $N_e\times 1$ vector denoting the a-priori channel energy information which is similar to correlating received energy samples with PDP. Averaged channel energy profile, $\mbs{\mathcal{E}}$, for IEEE 802.15.4a residential LOS channel model is as shown in Fig.~\ref{fig:PDP}. The performance of energy detector with WMESS algorithm is as shown in Fig.~\ref{fig:MSE_Vs_SNR_multi}. Even though, the WMESS estimation algorithm require more computation, its performance is better than matched filter at low SNRs for IEEE 802.15.4a CM1 multipath channel model. However, at high SNR there is a loss of performance compared to matched filter.

The performance degradation at high SNR with WMESS criteria for energy detector can become a problem when the TOA of the estimated target is dynamic in nature. In the next section, we will discuss a mechanism to fuse the estimates from multiple receivers to achieve improved performance for a dynamic TOA system. 

\section{Multi detector Fusion Using Kalman Filter}
\label{sec:Kalman}
When the estimated TOA of the target is dynamic in nature, then joint fusing and tracking using multiple independent TOA estimations from the  energy detectors can yield better performance. We employ Kalman structure as shown in Fig.~\ref{fig:KalmanStructure} to accomplish this.

The state of the system and its dynamic nature is represented by the state equation in the Kalman filter formulation, and is given by
\begin{equation}
  \label{eq:state}
  \mbf{s}[n]=\mbf{A}\mbf{s}[n-1]+ \mbf{u}[n],
\end{equation}
where $\mbf{s}[n]$, $\mbf{A}$, and $\mbf{u}[n]$ is given by
\begin{equation}
  \mbf{s}[n]=
   \begin{bmatrix*}[r] 
   \tau[n]\\
   \vtoa[n]
 \end{bmatrix*},  
 \label{eq:s1}
\end{equation}
\begin{equation}
  \mbf{A}[n]=
   \begin{bmatrix*}[r] 
     1 & 1 \\
     0 & \nu \\
 \end{bmatrix*},
 \label{eq:s2} \\
\end{equation}
\begin{equation}
  \mbf{u}[n]=
  \begin{bmatrix*}[r] 
     0 \\
     u_v[n]
 \end{bmatrix*},  
 \label{eq:s2}  
\end{equation}
where $u_v[n]$ is the noise with variance $\sigma_p^2$ and $\nu<1$ is a constant. The $\tau[n]$ is the TOA at n-th time interval and $\vtoa$ is an AR(1) processes with constant $\nu$, representing the dynamic nature of TOA. The plant noise $\mbf{u}[n]\sim \mathcal{N}\left(0,\mbf{\mathcal{Q}}\right)$, where $\mathcal{Q}$ is given by

  \begin{equation}
    \label{eq:q}
     \mathcal{Q}=
     \begin{bmatrix}0 & 0 \\ 0 & \sigma_p^2\end{bmatrix}.
  \end{equation}

\begin{figure*}[t]
  \centering
\fbox{\begin{minipage}{16cm}%
  \subfloat[MSE verses time ($\mb{SNR}=E_b/N_0=2~\mbox{dB}$)]{\includegraphics[scale=0.55]{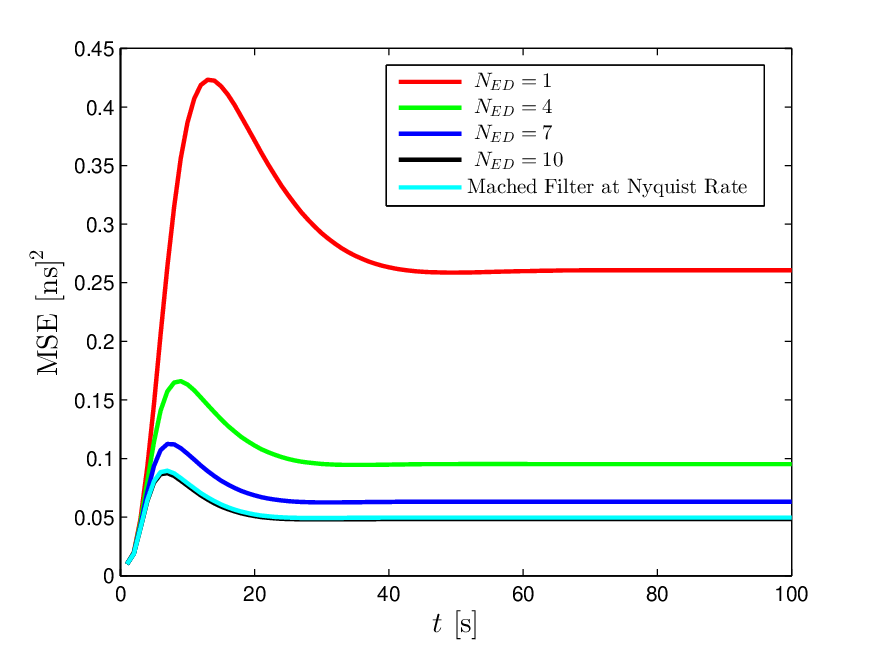}                 \label{fig:varianceVsD_SNR2}}
\subfloat[MSE verses time ($\mb{SNR}=E_b/N_0=6~\mbox{dB}$)]{\includegraphics[scale=0.55]{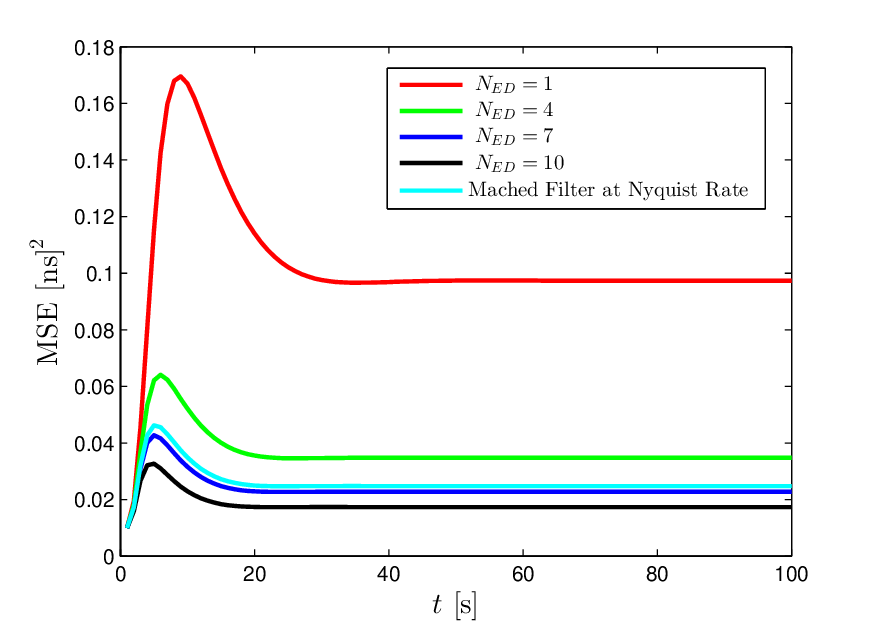} 
  \label{fig:varianceVsD_SNR6}}\\
\subfloat[MSE verses time($\mb{SNR}=E_b/N_0=10~\mbox{dB}$)]{\includegraphics[scale=0.55]{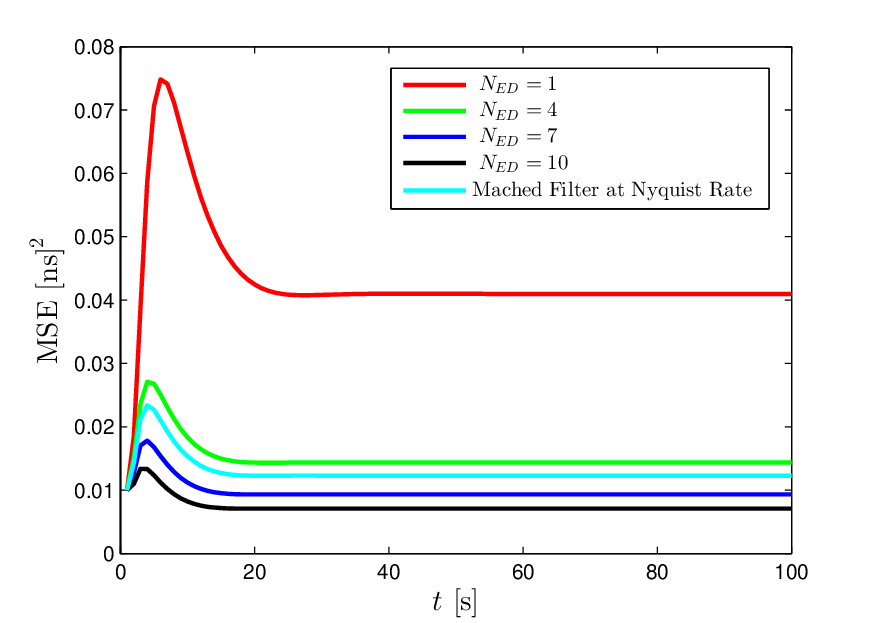} 
  \label{fig:varianceVsD_SNR10}}
\subfloat[MSE verses time ($\mb{SNR}=E_b/N_0=20~\mbox{dB}$)]{\includegraphics[scale=0.55]{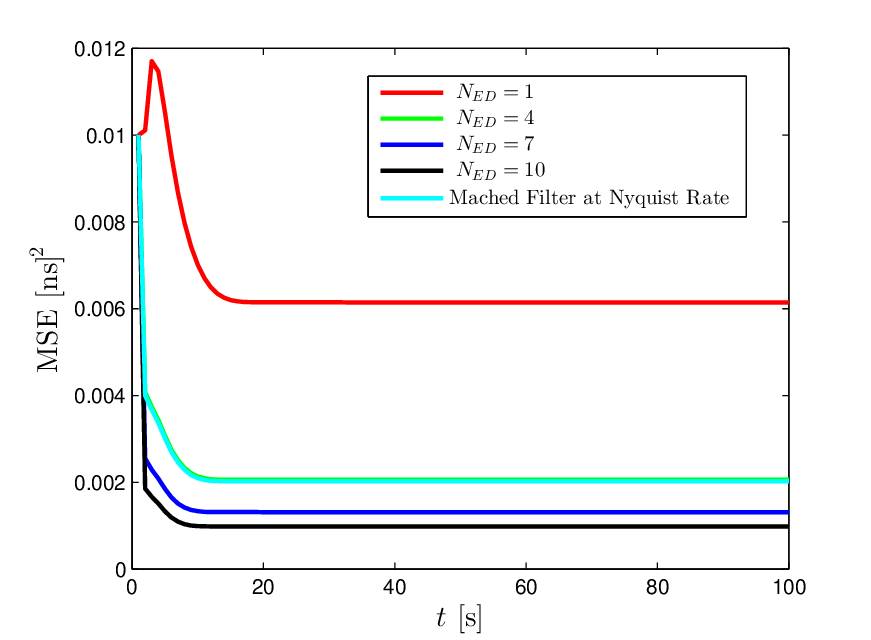} 
  \label{fig:varianceVsD_SNR20}}
\end{minipage}}
  \caption{Variation of MSE with time for digital matched filter and fusion of multiple sub-Nyquist energy detectors using Kalman filter at different SNRs}
\label{fig:var1}
\end{figure*}

The measurement matrix, $\mbf{y}=[\httoa^1,\httoa^2,\ldots,\httoa^N]$, is the independent TOA estimates from $N$ receivers. The measurement equation is given by
\begin{equation}
  \label{eq:meas}
  \mbf{y}[n]=\mbf{H}[n]\mbf{s}[n]+\mbf{w}[n],
\end{equation}
where 
\begin{equation}
  \label{eq:hmat}
  \mbf{H}[n]=
  \begin{bmatrix*}[r]
   1&0\\
   1&0\\
   \vdots & \vdots \\
   1&0\\
  \end{bmatrix*},  
\end{equation}
$\wn\sim\mathcal{N}(0,\mb{diag}(\sigma_1^2,\ldots\sigma_N^2))$. $N$ denotes the number of detectors and $\mb{diag}(\cdot)$ indicates the diagonal matrix with diagonal elements represented inside the brackets. 

The Kalman filter based fusion of  dynamic TOA from multiple energy detector estimates, $\mbf{y}[n]$, is as shown in Algorithm \ref{kalman}. At each iteration, the Kalman filter yields fused TOA estimate, $\httoa[n]$,  and its variance, $\hrtoa[n]$, from multiple energy detector estimates, $\mbf{y}[n]$. In our analysis, we have used same kind of detectors and also the reliability or importance  of each of the detectors are assumed same. By appropriately selecting the measurement matrix $\mbf{H}$ and covariance of the measurements, we can extend the Kalman filter formulation to accommodate different types of detectors with varying characteristics. In the next section, we will assess the performance of the proposed fusion and tracking method.

\section{Simulation Study}
\label{sec:ss}

\subsection{AWGN Channel}
\label{sec:awgn_sim}
AWGN channel is simulated with $F_s=8~\mb{GHz}$ with  $T_f=200~\mb{ns}$, $N_s=1$ and $B=4~\mb{GHz}$. We set $T_b=1~\mb{ns}$, thus resulting sampling rate of the energy detector is $1/8$-th the Nyquist rate.
Consider a linear variation of TOA ($\ttoa$) observed by a set of four independent energy detectors (refer Fig.~\ref{fig:KalmanStructure}). The observed TOA of the four detectors are as shown in Fig.~\ref{fig:4_detectors_observations_12dB} at $\mb{SNR}=12~\mb{dB}$. The tracked TOA using the Kalman formulation is shown in the Fig.~\ref{fig:4_detectors_tracked_12dB}. We consider prior state, covariance of prior state and plant noise variance equal to $[20,1]$, $0.01\mbf{I}_2$ and $\sigma_p^2=0.0001$ respectively. 

The mean square error (MSE) variation with time for the detection schemes with multiple energy detectors at various SNRs ($E_b/N_0$) are illustrated in the Fig.~\ref{fig:var1}. Notice that the steady state variance decreases as the number of energy detector estimates increases. Also, from  Fig.~\ref{fig:var1}, at high SNRs, only few energy detectors (operating at $1/8$-th rate of matched filter) are sufficient to achieve same performance as digital matched filter.

\begin{figure}[t]
  \centering
    \includegraphics[width=18pc]{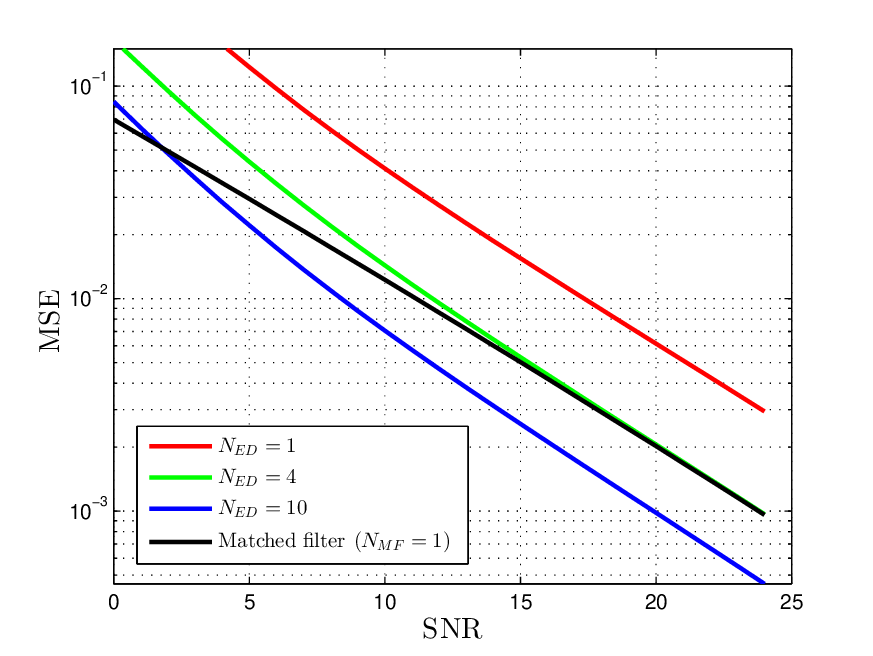}
  \caption{Steady state MSE verses time, for multiple energy detectors and digital matched filter.}
  \label{fig:steady}
\end{figure}

Figure~\ref{fig:steady}, shows the variation of steady state MSE with number of energy detectors (operating at $F_s/8$). Notice that steady state MSE of fusing $4$ independent energy detector estimates reaches that of digital matched filter estimate.

\begin{figure}[t]
  \centering
  \includegraphics[width=18pc]{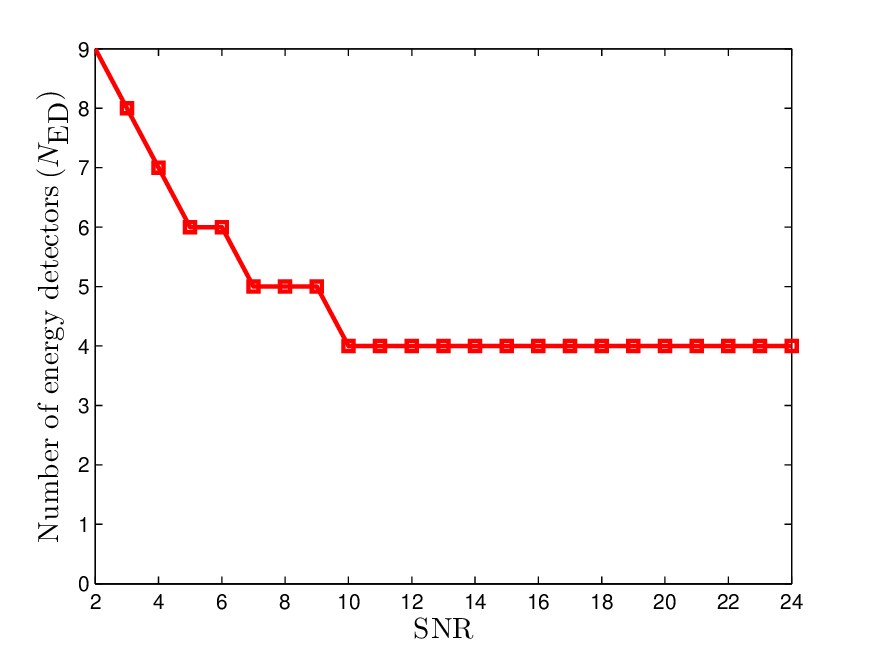}
  \caption{Number of sub-Nyquist energy detectors need to match the performance of matched filter operating at Nyquist rate.}
  \label{fig:var2}
\end{figure}

Figure~\ref{fig:var2}, shows the number of energy detectors operating in sub-Nyquist $(F_s/8)$ rate needed to match the performance of matched filter operating at Nyquist rate ($F_s$). At lower SNRs, more number of energy detectors, ($\NED$), are required to meet the matched filter performance. The number of energy detectors, $\NED$,  reduces with the increase of SNR and asymptotically  approaches to $4$, confirming with the analytical derivation of the previous section.


\subsection{Multipath Channel}
\label{sec:awgn_sim}
\begin{figure}[t]
  \centering
  \includegraphics[width=18pc]{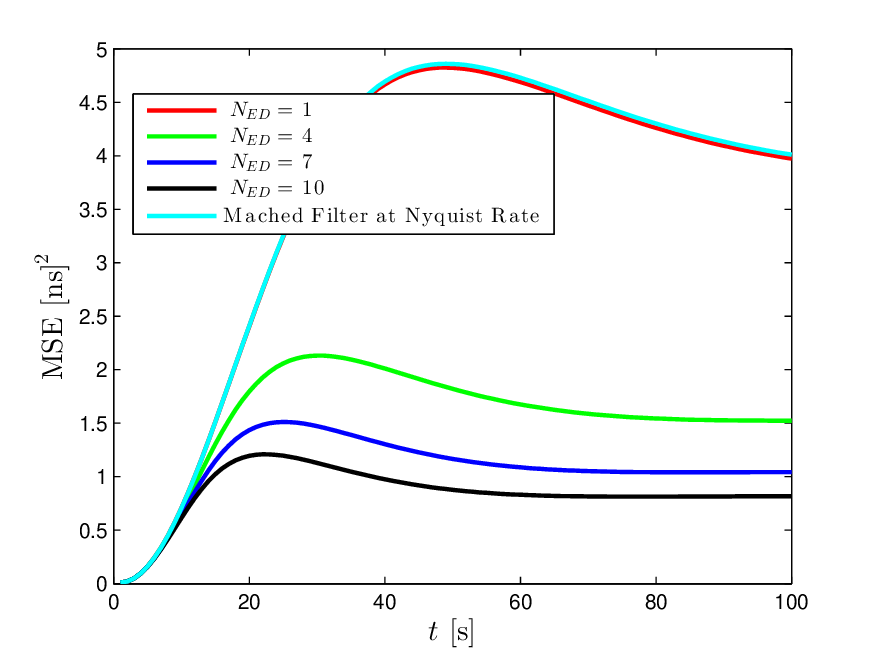}
  \caption{Variation of MSE for IEEE 802.14.4a CM1 channel model, with time for digital matched filter and fusion of multiple energy detectors (WMESS criteria is employed).}
  \label{fig:var1_multi}
\end{figure}

\begin{figure}[t]
  \centering
    \includegraphics[width=18pc]{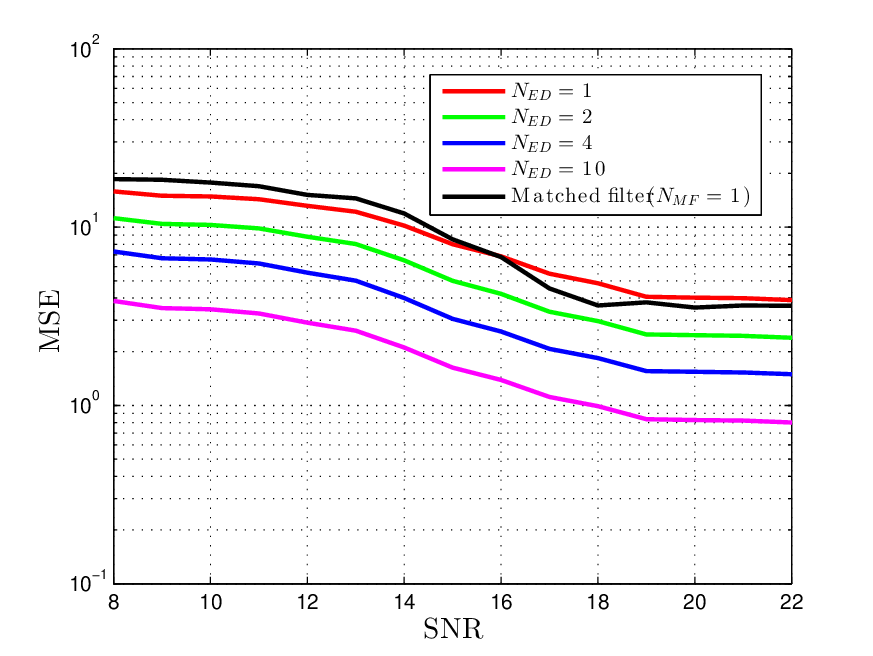}
  \caption{Steady state MSE verses time for IEEE 802.14.4a CM1 channel model, for multiple energy detectors (WMESS criteria is employed) and digital matched filter.}
  \label{fig:steady_multi}
\end{figure}

IEEE 801.15.4a CM1 channel model is simulated with $F_s=8~\mb{GHz}$ with  $T_f=200~\mb{ns}$, $N_s=1$ and $B=4~\mb{GHz}$. We set $T_b=1~\mb{ns}$, thus resulting sampling rate of the energy detector is $1/8$-th the Nyquist rate. We use WMESS algorithm discussed in the previous section to arrive at TOA estimates from the energy samples. We employ similar fusion and tracking strategy using the Kalman filter described in the previous section for the dynamic TOA model. The prior state information, covariance of prior state and other initialization parameters of the Kalman filter are kept same as in the previous section. Figure \ref{fig:var1_multi}, shows the variation of MSE with time at SNR of $20~\mb{dB}$ for different number of energy detectors. As expected the steady state variance decrease with the increase in the number of detectors. Figure \ref{fig:steady_multi}, shows the variation of steady state MSE with number of sub-Nyquist energy detectors, as discussed earlier a single energy detector with WMESS selection criteria outperforms the matched filter at low SNRs, however, at high SNRs more energy detectors are need to match the matched filter performance. Figure \ref{fig:steady_multi}, also illustrates that fusing two energy detector estimates ($\NED=2$)  can match the matched filer performance across the SNR region. 


\section{Conclusion and discussion}
\label{sec:Conclusion}
Estimating TOA with good accuracy is very important for localization and several other applications. UWB with its large bandwidth can yield  high precision ranging as evident from CRLB Equation \eqref{eq:crb}. For AWGN channel, matched filter based estimation yields optimum performance at high SNRs. However, matched filter requires Nyquist-rate sampling and is inefficient in terms of cost and power consumption. Energy detectors operating at sub-Nyquist rate is an interesting alternative as it can be designed using cost effective analog circuits and is power efficient, however, it lacks the precision in range measurements. In this paper, we showed that for AWGN channel, we can achieve the performance of a matched filter, by using a multiple sub-Nyquist energy detector estimates. We showed that number of energy detectors needed to meet the matched filters performance is high at low SNRs and reduces as SNR increases, and finally converges to $4$ as SNR increases asymptotically. This is analytically derived in \eqref{eq:neds2}, and confirmed through simulations (refer Fig.~\ref{fig:steady} and Fig.~\ref{fig:var2}). A Kalman filter with suitable choice of state-equation and measurement equations is designed to perform the dual task of tracking the TOA as well as fusing the multiple energy detector outputs. Filter equations are shown in Algorithm \ref{kalman} and performance in terms of MSE are demonstrated in Fig.~\ref{fig:var1} and Fig.~\ref{fig:steady}. Result indicate that for AWGN channel, the steady state variance drops with the increase of number of detectors and requires $4$ energy detectors to have the same performance as digital matched filter.

Proposed tracking and fusion strategy of energy detectors is analyzed in simulation for IEEE 802.15.4a CM1 multipath channel (residential LOS model) and results are demonstrated in Fig.~\ref{fig:MSE_Vs_SNR_multi}, Fig.~\ref{fig:var1_multi} and Fig.~\ref{fig:steady_multi}. Results indicate that the two sub-Nyquist sampled energy detector estimates with WMESS criteria can outperform digital matched filter at all SNR regions. 

We plan to further this work, by assessing the performance of the proposed fusion and tracking using multiple energy detectors using our  in-house UWB radio hardware discussed in Section \ref{sec:m}.

\bibliography{my}
\bibliographystyle{IEEE}
 
\end{document}